\documentclass[journal,comsoc]{IEEEtran}
\usepackage{booktabs} 
\usepackage{verbatim}
\usepackage{graphics} 
\usepackage{epsfig} 
\usepackage{epstopdf}
\usepackage{subfigure}
\usepackage{balance}
\usepackage{comment}
\usepackage{booktabs}
\usepackage{amsmath}
\usepackage{amssymb}
\usepackage{amsfonts}
\usepackage{bm}
\usepackage{colortbl}
\usepackage{tabularx}
\usepackage{color}
\usepackage{xspace}
\usepackage{graphicx}    
\usepackage{url}         
\usepackage{algorithm}
\usepackage{algpseudocode}
\usepackage{placeins}
\usepackage{multirow}
\usepackage{caption}
\usepackage{enumitem}
\usepackage{leading}
\usepackage{pgfplots}
\usepackage{tikz}
\usepackage{etex}
\usepackage{caption}
\usepackage{xcolor}
\usepackage{cite}
\usepackage{hyperref}

\ifCLASSINFOpdf
\else
\fi
\begin{document}
\title{A Novel Emergency Light Based Smart Building Solution: Design, Implementation and Use Cases}

\author{Weitao~Xu,
        Jin Zhang,
				Jun Young Kim,~\IEEEmembership{Member,~IEEE},
				Walter Huang,
				Salil S. Kanhere,
				Sanjay K. Jha,
		Wen Hu,
		Prasant Misra,~\IEEEmembership{Senior~Member,~IEEE}
\thanks{Weitao Xu is with Department of Computer Science, City University of Hong Kong. Email: weitaoxu@cityu.edu.hk. Salil S. Kanhere, Sanjay Jha and Wen Hu are with School of Computer Science and Engineering, University of New South Wales, Australia. Email:\{salil.kanhere,sanjay.jha,wen.hu\}@unsw.edu.au Jin Zhang, Jun Young Kim and Walter Huang are with WBS Technology, Australia. Email:\{jin.zhang,jun.kim,walter\}@wbstech.com.au
Prasant Misra is with TATA Consultancy Services Ltd., Email:prasant.misra@tcs.com}
}

\markboth{}%
{Xu \MakeLowercase{\textit{et al.}}}

\maketitle
\begin{abstract}
Deployment of  Internet of Things (IoT) in smart buildings has received considerable interest from both the academic community and commercial sectors. Unfortunately, widespread adoption of current smart building solutions is inhibited by the high costs associated with installation and maintenance. Moreover, different types of IoT devices from different manufacturers typically form distinct networks and data silos. There is a need to use a common backbone network that facilitates interoperability and seamless data exchange in a uniform way. In this paper, we present EMIoT, a novel solution for smart buildings that breaks these barriers by leveraging existing emergency lighting systems. In EMIoT, we embed a wireless LoRa module in each emergency light to turn them into wireless routers. EMIoT has been deployed in more than 50 buildings of different types in Sydney Australia and has been successfully running over two years. We present the design and implementation of EMIoT in this paper. Moreover, we use the deployment in a residential building as a use case to show the performance of EMIoT in real-world environments and share lessons learned. Finally, we discuss the advantages and disadvantages of EMIoT. This paper provides practical insights for IoT deployment in smart buildings for practitioners and solution providers.
\end{abstract}


%
\IEEEpeerreviewmaketitle
\section{Introduction}
\label{sec:introduction}
Buildings play a significant role in our modern society as they consume about $60\%$ of global energy and we spend 80--90\% of our lives in buildings (e.g., homes, offices)~\cite{energyconsumption}. With the advent of Internet of Things (IoT), the concept of building IoT (BIoT) is becoming increasingly popular nowadays. BIoT takes the idea of the IoT and applies it to commercial buildings. A BIoT system can connect all the smart devices (commonly known as things/objects) that are distributed across the building in an efficient and cost-effective manner. Furthermore, with big data and artificial intelligence (AI) technology a BIoT system can obtain valuable information from these data and immediately optimise and fully automate the buildings performance.  BIoT systems can improve utility of such buildings immensely such as reducing energy usage, facilitating rapid and efficient repair and maintenance. For example, the smart building deployed by Intel at Bangalore India leads to 20--30\% improvement in space utilisation and 10\% reduction in energy consumption every year~\cite{Intel}.

 While a significant number of infrastructures, technologies, and applications of BIoT have been developed, several barriers have prevented their widespread adoption such as heterogeneity of data, lack of ubiquitous connectivity and security. Among these challenges, the lack of low-cost backbone network that can achieve ubiquitous connectivity is the primary obstacle. Existing communication technologies have been successful in achieving ubiquitous connectivity but the solutions are expensive, lack flexibility and robustness. Traditionally, large BIoT networks have been a complex, fragmented system of different standards, devices, and services, making them extraordinarily hard to manage. Smart services and applications are delivered through a number of protocols, including Wi-Fi, Bluetooth Low Energy (BLE) and Zigbee, which presents a major challenge to different manufacturers and vendors. For example, a smart building may have a BLE network to support one or multiple BLE IoT apps and another Zigbee network to support one or more Zigbee IoT apps. To connect Wi-Fi enabled smart devices, we need to add another Wi-Fi network. The consequence is that we will have a complex set of networks that is challenging and costly to maintain and support. Unfortunately, most existing solutions are based on this strategy such as~\cite{stavropoulos2010system,7943156}. How to consolidate multiple heterogeneous networks into a single converged network in a low-cost way to simplify IoT endpoint onboarding and management remains an open problem.
 
 \begin{figure}[!t]
	\centering
		\includegraphics[width=0.8\linewidth]{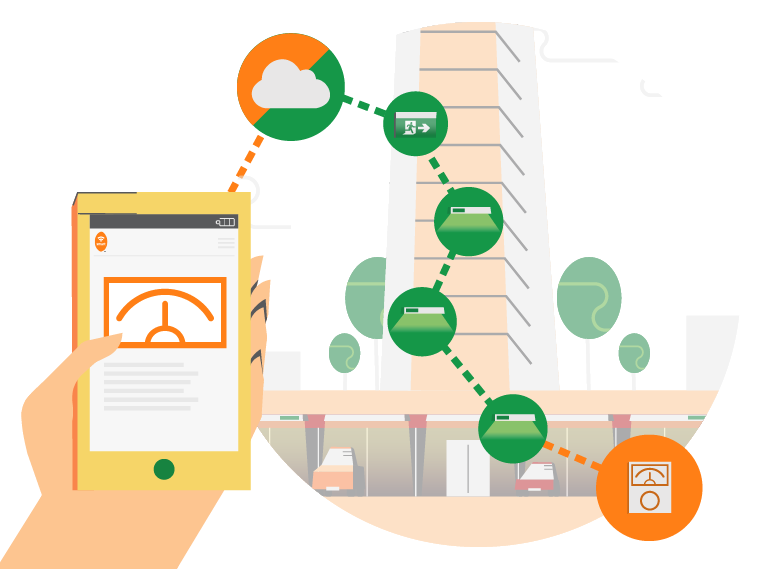}
	\caption{An overview of EMIoT.}
	\label{fig:diagram}
	\vspace{-0.2in}
\end{figure}

\begin{figure*}[!th]
	\centering
		\includegraphics[width=6in]{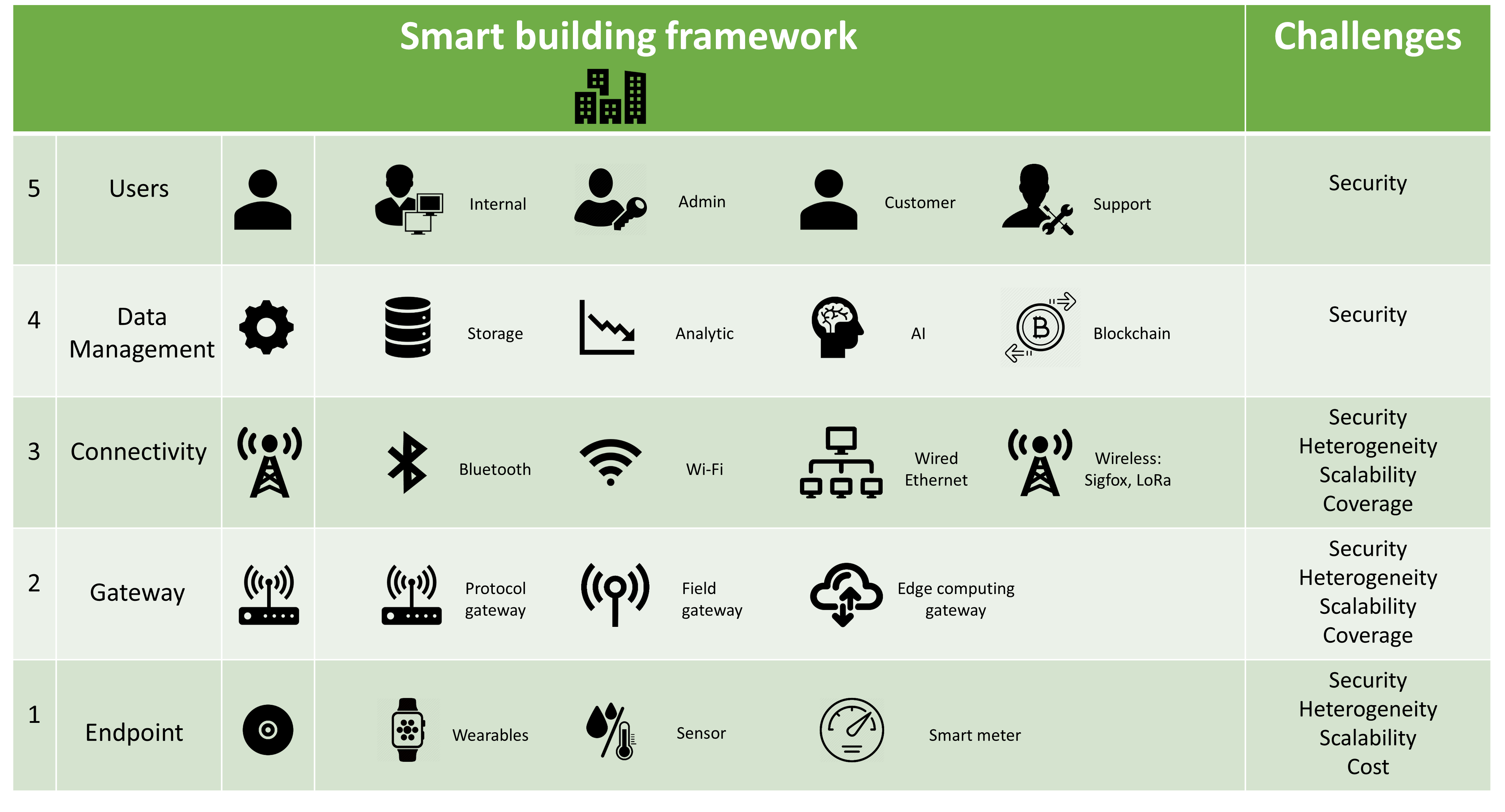}
	\caption{A conceptual BIoT framework .}
	\label{fig:architecture}
	\vspace{-0.2in}
\end{figure*}

This article aims to address this challenge by proposing a BIoT solution wherein, a common backbone network facilitates interconnectivity and seamless data exchange between different IoT devices and networks deployed within the building. Specifically, we present an emergency lighting IoT system for smart buildings---EMIoT (A demo of EMIoT can be found in~\cite{EMIoT}). The key feature of EMIoT is that it requires low CAPEX (capital expenditure) and is easy to deploy because it is built on top of a common facility in every buildings---emergency lights. In EMIoT, the ubiquitous emergency lighting system forms an interconnected backbone network with the help of LoRa communication technology. LoRa communication technology is one of the standards in Low-power Wide-area Network (LPWAN). Compared to other IoT communication technologies such as Wi-Fi and ZigBee, LPWAN features low-cost, low-power and wide coverage which make it highly suitable for BIoT smart objects whose requirements are low data rate, low energy consumption, and cost effectiveness. Built on the strength of LoRa and ubiquitous emergency lights, EMIoT presents a low cost and easy to deploy smart building solution that can provide a single converged network to connect any objects in a building. Fig.~\ref{fig:diagram} provides a brief overview of EMIoT.

We make two contributions in this article. First, we present a novel emergency light based smart building solution EMIoT. To the best of our knowledge, this is the first smart building solution that is built on top of emergency lights. Compared to current BIoT architectures, EMIoT presents several advantages as will be discussed later in this article. Second, EMIoT has been deployed in more than 50 buildings of different types such as residential building, office, and shopping centres in Sydney Australia and successfully running over two years. We share experience and lessons learned during deployment to provide practical insights for smart building solution providers.

The rest of the paper is organised as follows. Section~\ref{sec:IoTframework} presents a conceptual BIoT framework that can be used by researchers and engineers as a reference to design their own BIoT solutions. Section ~\ref{sec:currentchallenges} discusses the current challenges in BIoT. Then, we present the design details of EMIoT in Section~\ref{sec:architecture}. Next, we share our experience and lessons learnt in Section~\ref{sec:deployment}. Section~\ref{sec:benefits} discusses advantages and disadvantages of EMIoT and finally Section~\ref{sec:conclusion} concludes the paper.

\section{Conceptual BIoT Framework}
\label{sec:IoTframework}
Although the concept of BIoT has been around for decades,  a uniform and agreed BIoT reference framework and standard is lacking. In this paper, we present a conceptual BIoT framework that can be used as guide to design smart building systems (see Fig.~\ref{fig:architecture}).


\textbf{Layer 5: Users.} It is possible that a BIoT system has different types of users such as administrator, customer and system support. So this layer identifies the types of users who interact, either directly or indirectly, with the BIoT solution. This layer is used to assist service/solution vendors to define who their final clients are. Understanding all the users also enables an organisation to identify how their solution can meet different user requirements. 

\textbf{Layer 4: Data Management.} 
One of the most vital functions of a BIoT solution is the capability for an organisation to `manage' their devices and data through a diverse range of activities including: device connection management, data collection and processing, event tracking and handling, analytics, and interfaces to external systems. A BIoT system can generate a large volume of data such as temperature, energy usage and people's location. Traditional data storage is important but far from enough.
A smart BIoT system should be able to turn large amounts of data into useful, actionable information that can make the solution a truly smart system. The BIoT system can adopt some emerging technologies such as Artificial Intelligence (AI) and Blockchain to enhance the capability of the system.

\textbf{Layer 3: Connectivity.} Connectivity is an indispensable part of any BIoT solution. Devices can be connected via either wired communication or wireless communication technologies. A number of connectivity technologies have been invented to meet the diversified requirement of IoT, including: wired connectivity, Wireless Local Area Network (WLAN) such as Wi-Fi, Low Power Wide Area Network (LPWAN) such as LoRa, and Wireless Wide Area Networks (WWAN) such as 4G and 5G. Due to the heterogeneity nature of BIoT, there may exist several communication technologies in a single BIoT system. For example, BLE and Wi-Fi are often used to exchange data between personal devices such as smartphone and laptop. LPWAN is used in smart meters which do not require frequent communication. In some BIoT systems, 4G or 5G is used to connect a gateway to a remote server. Therefore, there is no clear difference between IoT and BIoT in terms of connectivity. This layer implements functionality for managing connected devices in a scalable and secure manner. Practitioners need to select proper communication protocols from different aspects such as power consumption, communication distance, bandwidth,  and capacity.

\textbf{Layer 2: Gateway.} A gateway is a bridge between end devices and users. It uploads data from an end device to a remote server for the user to monitor and access them. Meanwhile, the commands from the user are also transmitted to end devices via down-link. Similar to gateways in a general IoT framework, BIoT gateways depend on the type of connectivity technology and/or device types being used, since some devices will connect directly to the BIoT platform. Examples of BIoT gateways include protocol conversion gateway, field gateway,  and edge computing gateway where the latter can also perform certain functions defined by upper layers such as data analytic and AI.

\textbf{Layer 1: Endpoint.} In IoT, there is a vast variety of end devices ranging from personal devices such as mobile phones and laptops to public facilities such as street lights and connected cars. The endpoint in BIoT is a subset of IoT devices, and the commonly seen end devices include personal wearable devices, smart sensors (e.g., smoke detector), and smart meters (e.g., gas meter and water meter). These embedded devices posses the capability of monitoring and collecting data from ambient environment. The types of end devices in a BIoT system depends on the specific requirement of the solution.

\section{BIoT: Current Challenges}
\label{sec:currentchallenges}
Below, we summarise the challenges faced by current BIoT solutions.
 
 \begin{itemize}
    \item \textbf{Security.} Security is of utmost importance in BIoT. Ignoring security will lead to serious security issue and substantial economic loss.
    So, security should be the top priority of a BIoT solution. 
           
     \item \textbf{Heterogeneity.} In the context of smart building, the inherent heterogeneity of devices, networks and services makes efficient data management challenging. Moreover, the interoperability of devices and platforms is also important to facilitate seamless data exchange. To achieve this goal, we need a backbone network that can interconnect all heterogeneous devices such that the associated data can be accessed and exchanged seamlessly.
     
     \item \textbf{Scalability.} The assets and devices in a smart building are constantly evolving, and their number is also rapidly increasing. The BIoT solution, therefore, should have the capability to handle the growing number of devices and data.
    
    \item \textbf{Coverage.} BIoT requires a two-way communications network that can connect all the sensors, controllers and actuators in the building. However, traditional communication technologies either suffer from short communication range (e.g., ZigBee, Bluetooth) or require high power consumption (e.g., Wi-Fi, 4G, 5G). Thus, they are not suitable for future BIoT solutions where the smart sensors and objects are limited by their size, battery and processing ability. Moreover, the complex communication conditions in buildings due to concrete and steel structures and potential interference with the Wi-Fi networks installed by occupants also makes it challenging to achieve comprehensive coverage within the building, especially in basements and car parks.
     \item \textbf{Deployment/Maintenance Cost.} BIoT is traditionally expensive, complex, and requires customised installation, programming, and maintenance. Properly installed and connected IoT products can overcome the capital barriers of installing new devices and sensors. Therefore, a desirable BIoT solution should be cost-effective by using existing facilities or quick-to-install products.
 \end{itemize}
Because BIoT is a part of a wider IoT ecosystem, although the above challenges are specific to BIoT, they also apply to the general IoT system. Moreover, after we map the above challenges to the five layer BIoT framework, we have two findings. First, security is indeed the most important challenge because attackers can launch attack from each layer. For example, in Layer 3 connectivity, it is known that there are various types of attacks in a wireless network such as eavesdropping attack and Man-in-The-Middle (MITIM) attack. In Layer 1 endpoint, attackers can perform physical attack to breach the end device. Second, a low-cost ubiquitous and scalable backbone network is the key to solve the challenges from Layer 1 to 3. This is right the key feature of our EMIoT solution.
\begin{figure}[!t]
	\centering
		\includegraphics[width=2.8in]{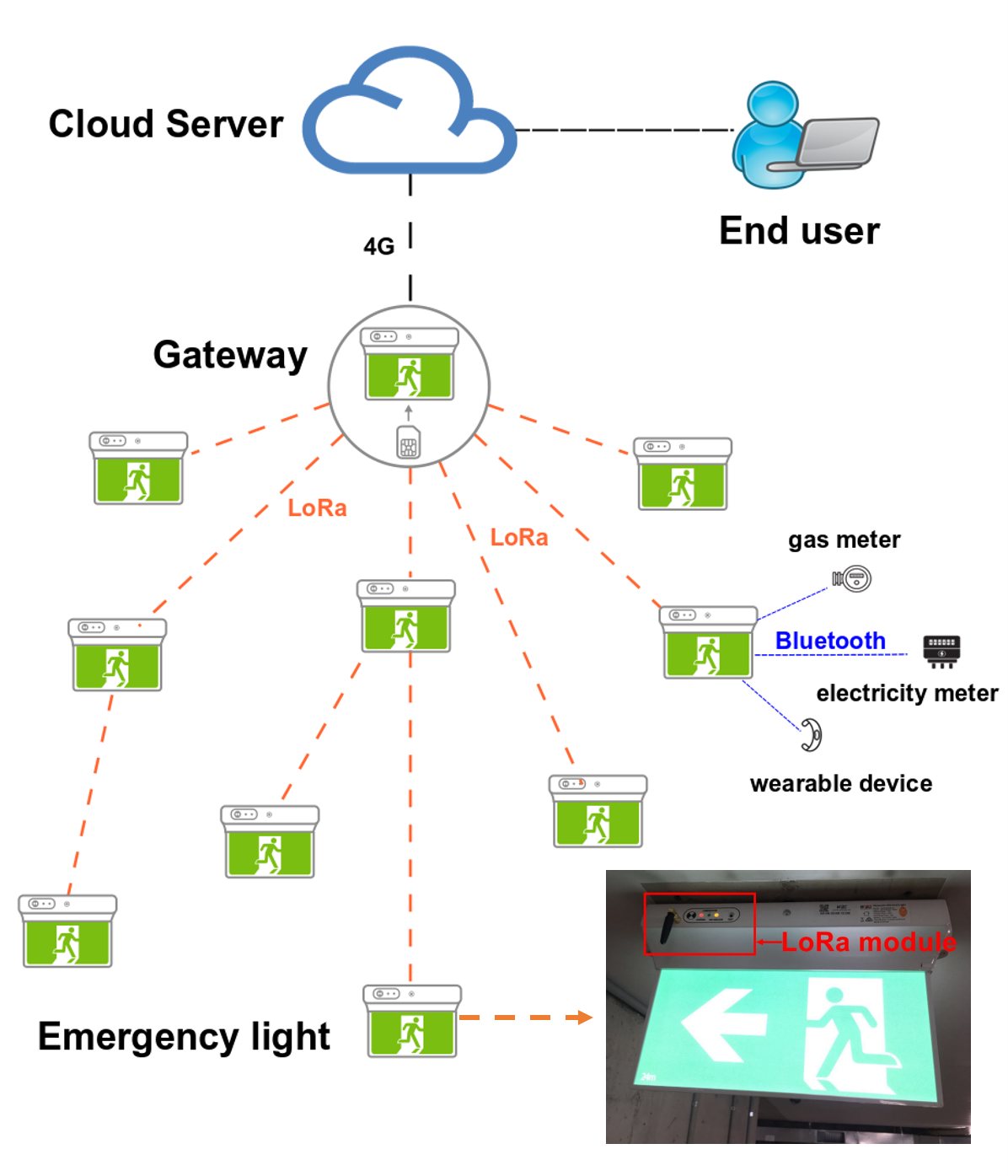}
	\caption{EMIoT architecture.}
	\label{fig:EMIoTarchitecture}
	\vspace{-0.2in}
\end{figure}

\section{EMIoT: A BIoT Barrier Breaking Solution}
\label{sec:architecture}

\subsection{System Design}
\label{subsec:design}
As mentioned above, one of the primary challenges in BIoT is the lack of a low-cost and ubiquitous backbone network. A backbone network is imperative for BIoT because it allows dynamic deployment of new networked electronic devices that help save energy and money along with greater environmental sustainability and building efficiencies. EMIoT addresses this challenge by building the BIoT system on top of emergency lights. The reason we adopt emergency lights is they are ubiquitous facilities in any buildings, thus providing an ideal basis for a low-cost and ubiquitous wireless backbone network. In this way, EMIoT enables numerous in-building smart objects to connect to a backbone network in a cost-effective way. Fig.~\ref{fig:EMIoTarchitecture} shows the system architecture of EMIoT. Each emergency light in the building can be turned into a smart node by installing our LoRa wireless module. Upon installation,  all the nodes form a mesh network autonomously and connect to a single gateway, one of the  mesh routers  configured with mobile connectivity to the Internet and  communicates with the cloud. Users (e.g., building manager) can remotely monitor and control the network via the smartphone app or a website.

\textbf{Cloud server.} 
EMIoT cloud leverages AWS located in Sydney Australia. The communication between the cloud and devices is managed by the underlying infrastructure. The cloud also provides user interface via the web or an application.

\textbf{Smart emergency lights.}
Normal emergency lights can be turned into smart modules with wireless connectivity by installing the EMIoT module. The EMIoT module consists of STM32 ARM-Cortex MCU and SemTech SX1272 radio chip (LoRa transceiver). The gateway uses the same hardware but it includes a 4G modem to relay the communication between the cloud and the network. The communication between the gateway and cloud uses a highly secure AWS IoT solution.


\textbf{LoRa mesh network.} EMIoT ensures widespread connectivity in a smart building by using LoRa communication technology. LoRa offers lower bandwidth, but has low power consumption and a higher penetration capability, which ensures the  coverage of the entire building. It can meet the low data transmission requirement of smart objects in the building such as smart meters. The wireless spectrum of LoRa uses sub Mega-hertz range, which is within the Industrial, Scientific and Medical (ISM) frequencies and thus free to use without any license fee. The default network structure of LoRa is a star network where all the nodes connect to a gateway directly. However, this is problematic in buildings as there are often blind spots due to  concrete walls/floors, interference from home Wi-Fi and equipments within buildings etc. Therefore, we design a novel LoRa mesh network protocol, where any LoRA node can connect to the same gateway via multiple hops. The designed LoRa mesh protocol is built on top of Low-Power and Lossy Networks (RPL) communication protocol~\cite{winter2012rpl}, where devices form a cooperative relaying network towards the most efficient network topology. RPL is based on IPv6 address and we optimise RPL in terms of duty cycle and parameters in transmission and routing to make it suitable for LoRa physical layer. The designed LoRa mesh network presents a self-healing capability that devices maintain the most efficient network topology according to the current environment.

\textbf{Over-The-Air (OTA) programming.} Wireless code dissemination is a fundamental function for IoT configuration management within a building as the IoT devices can be updated "in-situ" without requiring physical handling, thus saving manual labour costs and extended operational disruptions. In EMIoT, we designed a multicast-based OTA approach considering the low-bandwidth and the large number of devices. After a user uploads the new firmware image on the web UI, the gateway in the building receives the OTA command from the cloud. Then gateway multicasts a series of OTA packets to all the nodes in the LoRa network. After the propagation phase, devices request missing packets until they form a complete firmware image to update. A crucial feature of our OTA scheme is that it can complete software update without interrupting  regular operation of the emergency light while supporting seamless safety-critical features.

\textbf{Open network.} Once the EMIoT network is installed in the building premise, the system opens the backbone network to other smart devices such as water meters, smoke alarms, and power consumption monitoring nodes. Any device compatible with our protocol can join the network and have Internet connectivity resulting in a  truly smart building. The smart building can be a part of a large city/nation-wide IoT network. To enable the open smart building network, EMIoT leverages BLE communication. Any device with BLE capability can join EMIoT network and connect to the cloud. BLE modules are low price (around $\$1$) but feature powerful standardised communication. It is also possible to provide other link-layer connectivity for devices by adding corresponding communication modules (e.g., ZigBee, RFID) to our mesh nodes.

\textbf{Network security.} Since security is a crucial concern in the IoT field, EMIoT leverages AWS and  their MQ Telemetry Transport (MQTT) IoT services. AWS MQTT presents the state-of-the-art security features with the highest level of reliability such as device authentication, authorisation and data protection in the cloud. This guarantees the gateway-cloud connection security. For LoRa-Mesh security, we use highly secure methods such as Advanced Security Standard (AES) 128-bits encryption. The security of the open BIoT network is guaranteed by the BLE security standard.

\subsection{How EMIoT Addresses the Challenges}
\label{subsec:barrier}
Below, we briefly explain how EMIoT breaks the current barriers mentioned int he last section.
 \begin{itemize}
    \item \textbf{Coverage.}
Instead of installing new LoRa devices, EMIoT transforms existing emergency lights into meshed wireless modules to ensure connectivity across the building and thereby remove any “dark spots” or “dead zones”. With the strong penetrability of LoRa and ubiquity of emergency lights, EMIoT breaks the barrier of coverage, as no Access Points (APs) or cabling is required. 
    \item \textbf{Cost.} Since EMIoT uses emergency lights that already exist in all the buildings to double up and form a backbone network, the upfront investment cost is low. 
\item \textbf{Heterogeneity and Scalability.}
EMIoT extends connectivity to other IoT networks and devices through Bluetooth Low Energy (BLE). Any smart objects equipped with BLE modules in the building can easily connect to the EMIoT network via a nearby emergency light. The EMIoT network can support a large number of heterogeneous devices because each emergency light acts as a smart wireless router. Moreover, we design a LoRa-based mesh protocol for EMIoT, which can support large number of end devices. Therefore, the problem of heterogeneity and scalability is also solved. 
 \end{itemize}

\section{Use Case and Lessons Learnt}
\label{sec:deployment}
\subsection{Use Case}
\label{subsec:deployment}
So far, we have deployed the EMIoT system in more than 50 buildings of different types including residential buildings, shopping centres, warehouses and factories in Sydney Australia. In this paper, we use our deployment in a residential building located in Carlingford Sydney as a case study to show its performance. This building complex has three independent blocks---namely, A, B and C which share a three-level underground car park. The size of this building complex is about $45 \times 52 \times 22 ~m^3$ and it has ten levels in total. We deployed 69 emergency lights and two gateways (also an emergency light)in this building and tested the performance of EMIoT in six months' time. The floor plan and the locations of emergency lights are shown in Fig.~\ref{fig:floorplan}. 

\subsection{Results}
\begin{figure}
	\centering
		\includegraphics[width=3in]{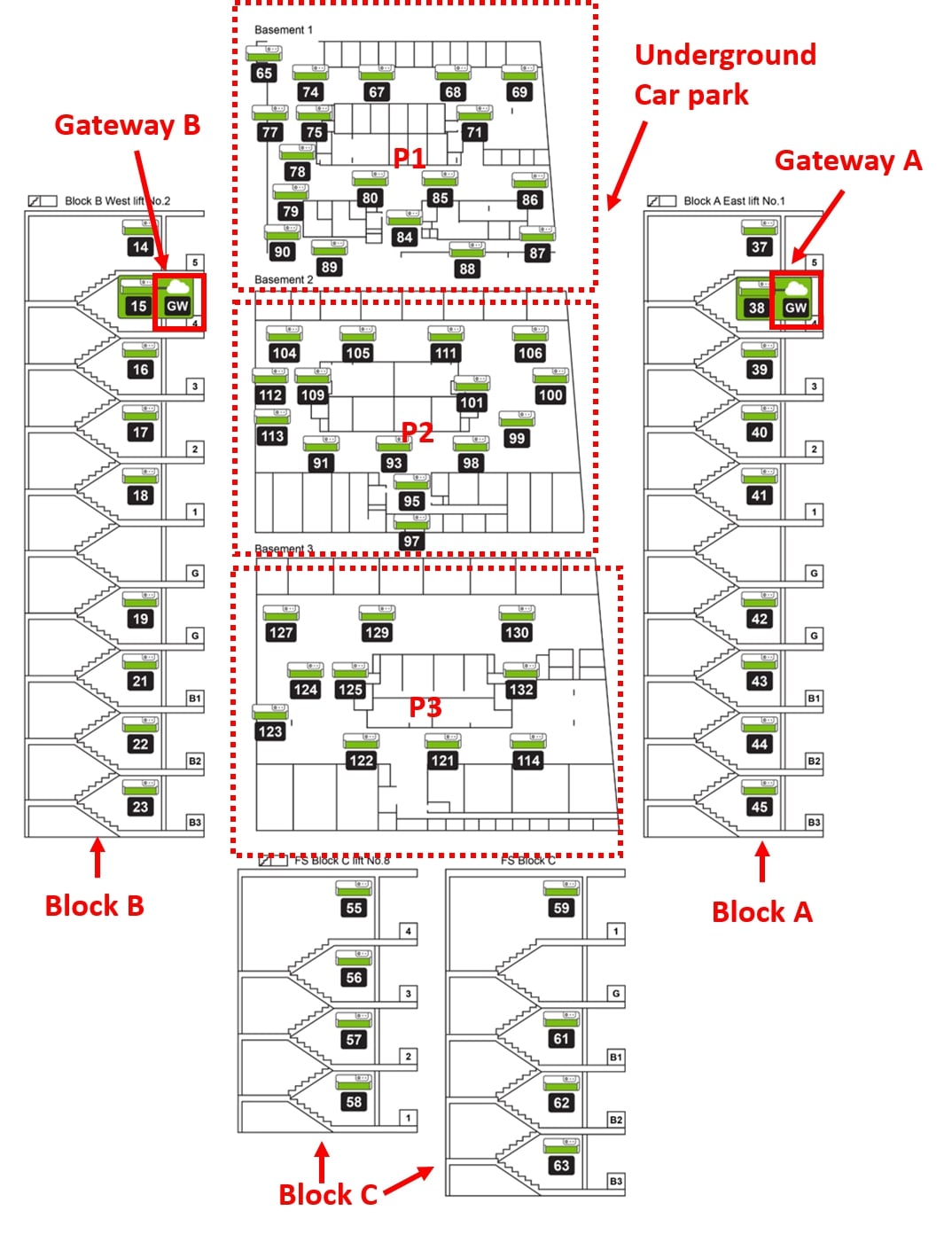}
		\caption{Floor plan.}
		\label{fig:floorplan}
		\vspace{-0.2in}
\end{figure}

\begin{figure}
	\centering
		\subfigure[Packet reception rate]{
		\includegraphics[width=2.2in]{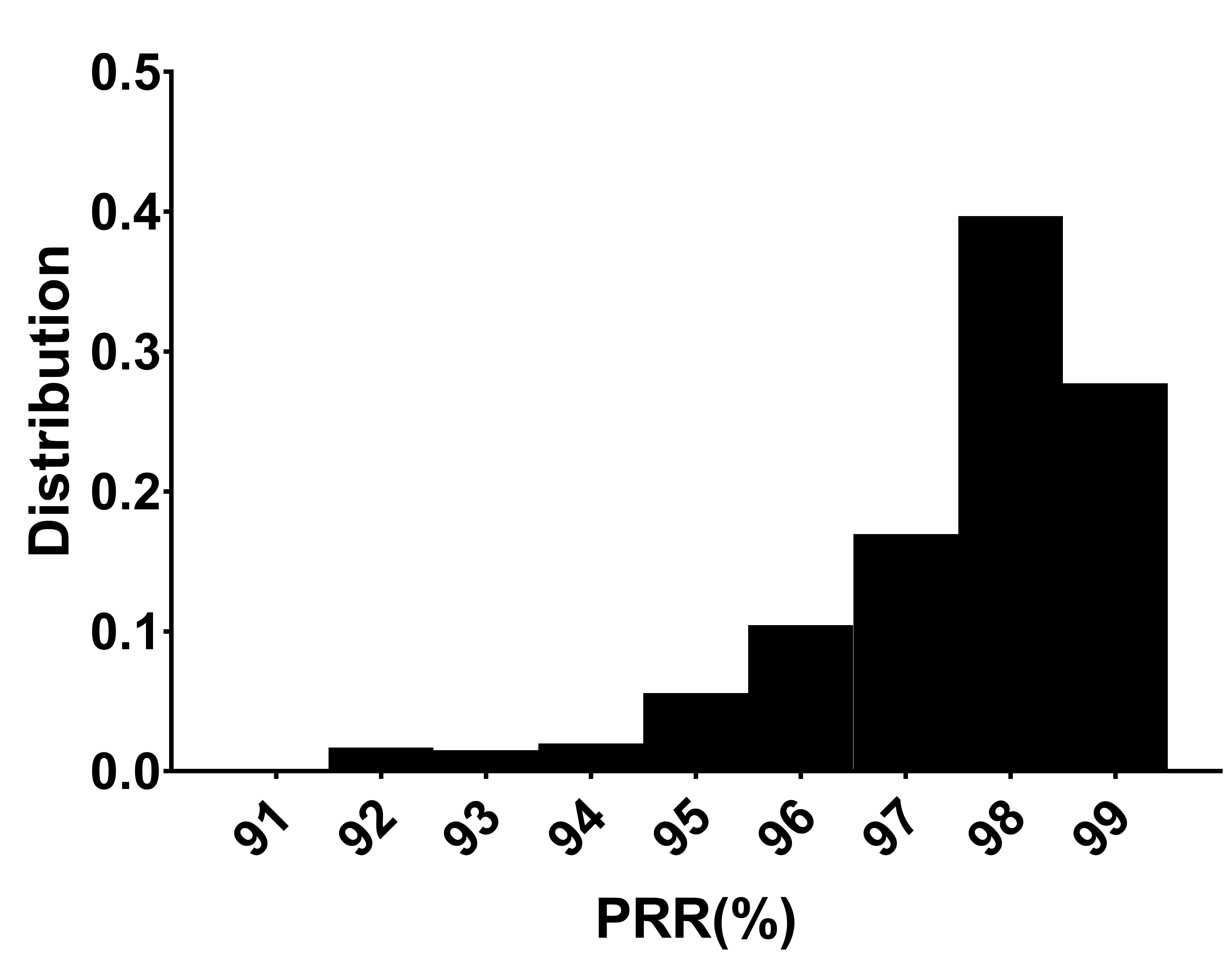}
		\label{fig:PRR}
		}
		\subfigure[Path length]{
		\includegraphics[width=2.2in]{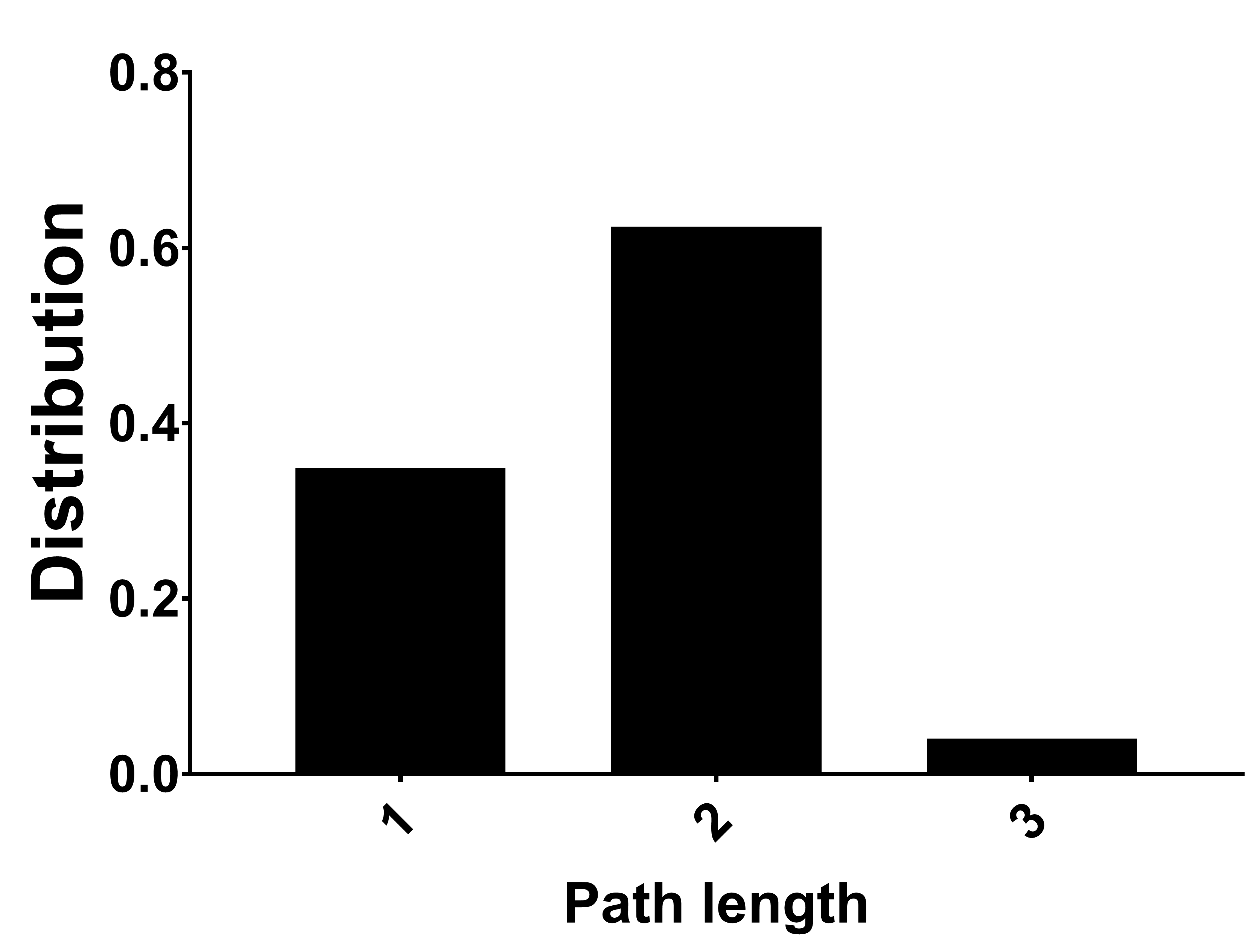}
		\label{fig:pathlength}
		}
	\caption{Evaluation results (using two gateways).}
	\label{fig:results}
	\vspace{-0.2in}
\end{figure}

We test the performance of EMIoT by using one gateway and two gateways, respectively. The locations of gateways are marked in Fig.~\ref{fig:floorplan}.  We find that one gateway can already cover all the devices in a building which demonstrate the feasibility of our LoRa mesh network. Each gateway is connected to our Amazon server via cellular network. We collected six months' data to analyse its performance. Overall, the network can achieve $97.8\%$ packet reception rate (PRR)  with one gateway and $98.6\%$ PRR with two gateways. Due to space limitation, we only plot the distribution of PRR using two gateways in Fig.~\ref{fig:PRR}. As shown in Fig.~\ref{fig:pathlength}, over $95\%$ of the emergency lights can be connected to the gateway within two hops, and the mean path length is 1.7 hops with two gateways. The system has been running successfully over two years which demonstrates its stability and robustness. We also test the performance of OTA and find that it takes about 7 hours to update a 125KB image for these emergency lights. If one emergency light runs into malfunction, the connected emergency lights can find links to gateway quickly which demonstrates the strong self-healing capability of our LoRa mesh network. The detailed performance of EMIoT can be found in our prior work~\cite{xu2019design}.

\subsection{Lessons}
During field study and deployment, the lessons learnt are detailed below.
\begin{itemize}
\item \textbf{Limitation of LoRaWAN.} The prominent communication protocol for LoRa is LoRaWAN which defines the MAC layer operations. LoRaWAN adopts star topology in which all the LoRa end devices are connected to a gateway. Then the gateway transmits data to a server using either Ethernet or wireless communication technologies such as 4G and Wi-Fi. However, in our field study, we found that if the gateway and LoRa emergency light are separated by more than seven floors then we see a remarkable decrease in packet delivery rate. The limitation of LoRaWAN is also discussed in other studies~\cite{bor2016lora,cattani2017experimental,adelantado2017understanding,liando2019known}. Therefore, the performance of LoRa is not as reliable as stated by the chip vendors. Additionally, in a one-hop network the location of the gateway should be carefully considered. A common practice is to install the gateway on the rooftop so it has better connectivity to remote server. This manner, however, will reduce the link quality between gateway and end devices installed in the building. Therefore, there is a trade-off between gateway-server connectivity and gateway-end device connectivity. EMIoT do not have such a problem because it uses LoRa mesh network: if a node cannot connect with gateway directly, it can find a link to gateway via multiple hops. So we can put gateway anywhere as long as it has stable connections to the cloud. 

\item \textbf{Complexity of environment.} Basements and car parks are known to be challenging environments for wireless communication due to high path loss and dynamic channel conditions~\cite{silva2009empirical,akyildiz2009signal}. We found that the path loss changes greatly even when we move a short distance on the same level. Our previous studies also revealed that the packet delivery rate varies remarkably at different locations of the same level of an underground parking~\cite{xu2019design,xu2019measurement}. Consequently, most devices in the underground parking require three hops to reach the gateway, which means only mesh topology can cover the basement and car park properly. Another observation is that even buildings with similar layout lead to different network typologies. We believe it is mainly due to the differences in building materials. Another finding is that the walls offer better LoRa signal penetration capability than floors. This is shown in our topology analysis in~\cite{xu2019design} that the parents of some devices are located in another block rather than devices on the upper and lower floors even though they may be physically closer.
\end{itemize}

\section{Discussion}
\label{sec:benefits}
There is no single definition for a “smart” building and different buildings have distinct requirements. Some buildings aims to reduce energy consumption and upfront cost, while others may strive to improve resident satisfaction and operational efficiency. For example, in 2016, Intel created its first IoT-enabled smart building in Bangalore, India with the aim to reducing resource usage, improving operational efficiency, and increasing occupant comfort~\cite{Intel}. Different requirements leads to different design principles and system structures. Moreover, the definition and functionalities of a smart building will also be continuously evolving with time.  Clearly, EMIoT is not a solution that can meet the requirements of any BIoT system. With this in mind, we discuss the advantages and disadvantages of EMIoT in this section.

The goal of EMIoT is to provide a uniform backbone network for low-power and low data rate objects in a smart building in a low cost way. The benefits of EMIoT are detailed below.
\begin{itemize}
\item \textbf{Uniform Backbone Network.} Many buildings have a combination of proprietary systems that do not “talk” to each other. EMIoT addresses this problem by building an uniform backbone network that is ubiquitous in the whole building. Any third party sensors, actuators and meters can be connected to EMIoT network. 
\item \textbf{Low Cost.} The installation/maintenance cost can be greatly reduced, as existing emergency lights with power supply are utilised as backbone by replacing the control module. Even though cable-based systems provide reliable and high bandwidth, they require cable installation and maintenance, which costs a lot of effort and investment. In contrast, our system uses meshed LoRa wireless communication technology which features low power communication and strong penetration capability. No additional equipment is necessary in the building as the gateway possesses independent 4G mobile connection. 
\item \textbf{Full Building Coverage.} As emergency lights are ubiquitous in any building by law for safety, they are widely distributed. Thus, EMIoT can provide the full network coverage to realise ubiquitous connectivity.

\item \textbf{Environmental friendly.} Eco-friendly smart building solutions are imperative as buildings consumes more than 50$\%$ of the electricity load in the world. In Australia, up to $50\%$ of CO2 emissions is attributed to energy consumption of buildings. EMIoT enables large-scale energy-output monitoring and intelligent actuation in smart buildings which will reduce energy consumption and greenhouse gas emissions significantly.
\end{itemize}

In EMIoT, the ubiquitous communication is achieved by ubiquitous emergency lights and LoRa which has strong penetration communication ability. The emergency lights can be replaced by other mandatory safety equipment installed in buildings such as smoke detectors. The other LPWAN communication technologies such as NB-IoT can also be used to replace LoRa. But in our solution, the advantage of LoRa over NB-IoT is that we can set up and manage our own network (i.e., one gateway for one building) while NB-IoT is dependent on cellular network coverage.

Despite the advantages above, EMIoT is not a mast key to any smart building solution. The application of EMIoT is limited by its low data rate. The communication technology used in EMIoT is LoRa  which is considered to be the future wireless communication standard for IoT. However, the long range of LoRa is achieved by sacrificing bandwidth and data rate. Compared to legacy communication technologies such as Wi-Fi and ZigBee whose bandwidth is in the order of Kbps and Mbps, the bandwidth of LoRa only varies from several hundreds of bps to several Kbps depending on parameter setting. The limited bandwidth available in LoRa makes it unsuitable for high data rate applications such as camera surveillance and file transferring. Therefore, EMIoT is only suitable for application scenarios that require low data output such as smart meters.



\section{Conclusion}
\label{sec:conclusion}
BIoT is an an important component of future  smart building systems since it is closely related to people's daily lives. In this paper, we present a novel IoT networking solution named EMIoT. EMIoT is built on top of ubiquitous emergency lights for cost effectiveness and ease of deployment. With our novel LoRa mesh network, EMIoT achieves ubiquitous connectivity that can server as a backbone network for smart objects in the building. Our deployment in Sydney serves as an example to demonstrate its performance over two years. We also share some hands on experience from our deployment. Currently, EMIoT has been deployed in more than 50 buildings in Sydney and we are working closely with our partners to bring it into many new types of buildings. We hope that our work can provide new ideas, insights and inspirations for practitioners to facilitate the development of BIoT.

\ifCLASSOPTIONcaptionsoff
  \newpage
\fi

\section*{Acknowledgements}
This research was supported by Australian Research Council Linkage Project LP160101260.


\bibliographystyle{IEEEtran}
\bibliography{sample}

\begin{IEEEbiographynophoto}{Weitao Xu} (GS'16-M'17) is currently an Assistant Professor at the Department of Computer Science, City University of Hong Kong. He received his PhD degree from the University of Queensland in 2017. 
\end{IEEEbiographynophoto}
\vspace{-0.4in}
\begin{IEEEbiographynophoto}{Jin Zhang} (GS'14) is currently a postdoc at Shenzhen Institutes of Advanced Technology Chinese Academy of Sciences. He received PhD from the University of New South Wales (UNSW) in 2017. 
\end{IEEEbiographynophoto}
\vspace{-0.4in}
\begin{IEEEbiographynophoto}{Jun Young Kim} (GS'14) received PhD from the University of New South Wales (UNSW) in 2017. His main research consists of security and management of the Internet of Things (IoT). He is now leading the R$\&$D team at the WBS technology in Sydney to develop a smart building service.
\end{IEEEbiographynophoto}
\vspace{-0.4in}
\begin{IEEEbiographynophoto}{Walter Huang} is the director of WBS Technology Australia. WBS Technology is recognised as Australia's leading manufacturer of high quality LED Emergency Lighting, LED Exit Signs and LED Lighting Systems.
\end{IEEEbiographynophoto}
\vspace{-0.4in}
\begin{IEEEbiographynophoto}{Salil S. Kanhere} is currently a Professor with the School of Computer Science and Engineering, University of New South Wales. His current research interests include pervasive computing, crowdsourcing, and sensor networks.
\end{IEEEbiographynophoto}
\vspace{-0.4in}
\begin{IEEEbiographynophoto}{Sanjay K. Jha} (SM'08) is a Professor, head of NetSyS and director of CySPri Laboratory at the School of Computer Science and Engineering at the University of New South Wales. 
\end{IEEEbiographynophoto}
\vspace{-0.4in}
\begin{IEEEbiographynophoto}{Wen Hu} (S'04–M'06–SM'12) is an Associate Professor with the School of Computer Science and Engineering, the University of New South Wales. His research interests include sensor networks, IoT and low power communications. 
\end{IEEEbiographynophoto}
\vspace{-0.4in}
\begin{IEEEbiographynophoto}{Prasant Misra} is a scientist with TCS Research \& Innovation, where he works on intelligent cyber-physical systems for smart mobility. His research experience spans across various aspects of mobile sensing and computing, with a current focus on decision sciences. 
\end{IEEEbiographynophoto}
\end{document}